\documentclass[12pt, 
]{article}
\usepackage{booktabs}

\usepackage[utf8]{inputenc}
\usepackage{amssymb}
\usepackage{amsmath}

\usepackage{geometry}
\newtheorem{theorem}{Theorem}

\newtheorem{definition}{Definition}
\newtheorem{remark}{Remark}

\newenvironment{proof}[1][Proof]{\emph{#1.} }{\  \hfill $\square $ \vspace{5 pt}}

\usepackage{natbib}
\usepackage{amssymb}
\usepackage[dvips]{graphics}
\usepackage{amsmath}

\usepackage{mathpazo}
\usepackage{enumerate}

\usepackage{amsfonts}

\usepackage{amssymb}

\usepackage{enumerate}

\usepackage{mathrsfs}

\usepackage[usenames]{color}
\usepackage{cancel}

\usepackage{pgf,tikz}

\usetikzlibrary{decorations.pathreplacing,angles,quotes}
\usetikzlibrary{arrows.meta}
\usetikzlibrary{arrows, decorations.markings}
\tikzset{myptr/.style={decoration={markings,mark=at position 1 with %
       {\arrow[scale=2,>=stealth]{>}}},postaction={decorate}}}

\usepackage{hyperref}
\hypersetup{
    colorlinks,
    citecolor=blue,
    filecolor=black,
    linkcolor=blue,
    urlcolor=blue
}

\usepackage{pgf,tikz}
\usetikzlibrary{arrows}

\usepackage{fancyhdr}

\usepackage{soul}

\usepackage{emptypage}

\newcommand*\samethanks[1][\value{footnote}]{\footnotemark[#1]}

\usepackage[T1]{fontenc}
\DeclareFontFamily{T1}{calligra}{}
\DeclareFontShape{T1}{calligra}{m}{n}{<->s*[1.44]callig15}{}
\DeclareMathAlphabet\mathcalligra   {T1}{calligra} {m} {n}

\usepackage{multirow}
\usepackage{array}
\usepackage{mathtools}
\usepackage{arydshln}
\usepackage{threeparttable}
\usepackage{booktabs,caption}

\usepackage{ifthen}
\newboolean{showcomments}
\setboolean{showcomments}{true}

\newcommand{\pablo}[1]{  \ifthenelse{\boolean{showcomments}}
{\textcolor{green!50!black}{(T: #1)}}{}}
\newcommand{\marcelo}[1]{\ifthenelse{\boolean{showcomments}}
{\textcolor{red}{(M: #1)}}{}}
\newcommand{\agustin}[1]{  \ifthenelse{\boolean{showcomments}}
{\textcolor{blue!50!black}{(T: #1)}}{}}

\begin{document}

\title{ A Note on Obvious Manipulations of Quantile Stable Mechanisms%
\thanks{%
We acknowledge the financial support
from UNSL through grants 032016, 030120, from CONICET through grant
PIP 112-200801-00655, and from Agencia Nacional de Promoción Cient\'ifica y Tecnológica through grant PICT 2017-2355.}}


\author{R. Pablo Arribillaga\thanks{
Instituto de Matem\'atica Aplicada San Luis (CONICET-UNSL) and Departamento de Matemática, Universidad Nacional de San Luis, San Luis, Argentina, and RedNIE. Emails: \href{mailto:rarribi@unsl.edu.ar}{rarribi@unsl.edu.ar} 
and \href{mailto:ebpepa@unsl.edu.ar@unsl.edu.ar}{ebpepa@unsl.edu.ar}
} \and Eliana Pepa Risma\samethanks[2] 
}
\date{\today}

\maketitle
\begin{abstract}
In two-sided matching markets with contracts, quantile (or generalized median) stable mechanisms represent an interesting class that produces stable allocations which can be viewed as compromises between both sides of the market. These mechanisms balance the competing priorities of the parties while maintaining stability.  This paper explores obvious manipulations of quantile stable mechanisms. Unfortunately, we get that any quantile stable mechanism different to the doctor-proposal DA is obviously manipulable. Our findings highlight the trade-offs between robustness to manipulation and other desirable properties, such as equity, in the design of stable matching mechanisms.

\bigskip

\noindent \emph{JEL classification:} D71, D72. \bigskip

\noindent \emph{Keywords:} obvious manipulations, matching, contracts, mechanism design, quantile stable

\end{abstract}

\section{Introduction}

A two-sided matching model with contracts involves a market comprising two distinct groups, such as doctors and hospitals, where contracts represent agreements between specific pairs of agents. In one-to-one models, each doctor or hospital can sign at most one contract, and agents have strict preferences over the contracts involving them.\footnote{For clarity, we present our (negative) result in a one-to-one context; however, it is straightforward to see that the result also applies to the more general many-to-one (with contracts) framework.
Most of the cited papers focus on the many-to-one framework, often incorporating assumptions about the preferences of the many-side, such as responsiveness, substitutability and/or law of aggregate demand. Naturally, their positive results hold in the one-to-one model.} A set of contracts satisfying this constraint is called an \emph{allocation}. 

An allocation is \textit{stable} if no doctor-hospital pair prefers to firm a contract outside the allocation. \cite{hatfield2005matching} established that the set of stable allocations is non-empty and that the optimal stable allocation for each side of the market can be computed using generalizations of the Deferred Acceptance (DA) algorithm by \cite{gale1962college}. 

Among stable allocations, the \emph{quantile stable} ones have gained attention as compromises between both sides of the market \citep[see, e.g.,][]{teo1998geometry, schwarz2009median, chen2014ranking, chen2016manipulability, chen2016median, chen2021quantile, fernandez2020deferred}. 
To present such allocations, suppose that there are $k$ stable allocations in a given market, and let $j\in \{1,...,k\}$ be arbitrary. For each doctor, consider their $j$-th best stable contract. Now, consider the set of contracts formed by selecting each doctor's $j$-th best stable outcome. \cite{chen2016median} showed that this set of contracts constitutes a well-defined stable allocation for each $j\in \{1,...,k\}$. \footnote{For the classical one-to-one model without contracts, \cite{teo1998geometry, schwarz2009median} proved the existence of quantile stable matchings in one-to-one matching.}
 They are called the\textit{ quantile stable allocations}. The extremal quantile stable allocations ($j=1$ and $j=k$) coincide with the optimal stable allocations for each side of the market. Furthermore, as $j$ moves from $1$ to $k$ quantile stable allocations represent different "levels" of compromise between both sides of the market. In particular, the \textit{median stable allocation} (which corresponds to $j=\frac{k}{2}$ or $j=\frac{k+1}{2}$ depending on whether $k$ is even or odd), is a notable option among the non-extremal quantile stable outcomes.
 Market designers consider it an appealing solution due to its symmetric treatment of both sides of the market. This perspective is supported by experiments from \cite{echenique2024experimental}, which showed that in decentralized matching settings, agents tend to coordinate on the median-stable allocation when it exists. Quantile stable allocations are also called \emph{generalized median-stable} \textit{allocations} in \cite{chen2016median}.

In our analysis, we assume that doctors are strategic and have private preferences, while hospitals' preferences are fixed and publicly known, following standard assumptions in the literature \citep[e.g.,][]{hirata2017stable, iwase2022equivalence}. A \emph{(matching) mechanism} is a mapping that for each preference profile of doctors, selects an allocation. In particular, we focus on \emph{quantile stable mechanisms} \citep[see][]{  
chen2014ranking,chen2016manipulability, chen2021quantile} which select a quantile stable allocation maintaining a given ``level'' of compromise between doctors and hospitals in each market.
 
The quantile stable mechanism that always selects the optimal stable allocation for doctors corresponds to the doctor-proposing DA, which is well known to be strategy-proof \citep{dubins1981machiavelli, roth1982economics}. However, any other quantile stable mechanism may be susceptible to manipulation.

Given that manipulations cannot be entirely avoided in most quantile stable mechanisms, we explore whether they prevent obvious manipulations as defined by \cite{troyan2020obvious}. A manipulation is obvious if it is significantly easier for agents to identify and execute, based solely on their understanding of the possible outcomes. While \cite{troyan2020obvious} demonstrated that all (quantile) stable mechanisms are not obviously manipulable in a model without contracts, we show that this result does not hold in presence of contracts.  In fact, our main theorem in this paper shows that the doctor-proposing DA is the only quantile stable mechanism that remains not obviously manipulable in the considered context. 

The present work complements our earlier results in \cite{arribillaga2025obvious}, where we analyzed stable-dominating mechanisms in many-to-one models with and without contracts. In that study, we showed that, while the doctor-proposing DA is not obviously manipulable, other mechanisms such as the hospital-proposing DA become susceptible to obvious manipulations in presence of contracts. 

Recent studies have extended the notion of obvious manipulation to other settings, such as one-sided matching markets \citep{troyan2020essentially}, voting \citep{aziz2021obvious, arribillaga2024obvious}, cake-cutting \citep{ortega2022obvious}, and allocation problems \citep{psomas2022fair, arribillaga2025not}.

The remainder of this paper is organized as follows. Section \ref{section prelim} introduces the model and key concepts. Section \ref{section main} presents our result.

\section{Model}\label{section prelim}

In the one-to-one matching model with contracts, there is a finite universal set of contracts $\mathbf{X}$, and two disjoint finite sides: a set of doctors $D$ and a set of hospitals $H$. Each contract $x \in \mathbf{X}$ is bilateral, involving exactly one doctor $x_{D} \in D$ and one hospital $x_{H} \in H$. The set $\mathbf{X}$ may include two or more contracts involving the same pair of agents $(d,h) \in \mathbf{D \times H}$. 
A possible solution to our assignment problem, referred to as 
\textbf{allocation}, can be characterized as follows: $Y\subseteq \mathbf{X}$ is an allocation if  $x\neq y$ implies $x_{D}\neq y_{D}$ and $x_{H}\neq y_{H}$, for all $x,y\in Y$. Let $A(\mathbf{X})$ denote the set of all allocations.
Given $Y\in A(\mathbf{X})$ and an agent $i \in D \cup H$, we denote by $Y_{i}$ the only contract in $Y$ involving $i$, if any. Otherwise, $Y_{i}=\emptyset$. 

Each agent $i \in D \cup H$ has preferences over $\mathbf{X}_{i}\cup \{\emptyset\}$.\footnote{Preferences are antisymmetric, transitive, and complete binary relations on $\mathbf{X}_{i}\cup \{\emptyset\}$. }  In our analysis, as is common in the literature, we assume that only one side of the market —the doctors— has unknown preferences and could have a strategic behavior, while hospitals' preferences are fixed and commonly known.
An arbitrary preference for doctor $d$ will be denoted by $P_{d}$, and the associated weak preference relation will be represented by $R_{d}$.\footnote{That is, for all $x, y \in \mathbf{X}_{i}$, we have $xR_{d}y$ if and only if either $x = y$ or $xP_{d}y$.} The set of all possible preference relations for a doctor $d$ in a given market will be denoted by $\mathcal{P}_{d}$. A preference profile $P = (P_{d})_{d \in D}$ specifies a preference relation for each doctor, and the set of all possible preference profiles in the market is represented by $\mathcal{P} = \prod_{d \in D} \mathcal{P}_{d}$. Finally, for each profile $P$ and doctor $d \in D$, we will denote by $P_{-d}$ the subprofile in $\mathcal{P}_{-d} = \prod_{i \in D \setminus \{d\}} \mathcal{P}_{i}$ obtained by removing $P_{d}$ from $P$.

\begin{definition}
A \textbf{(matching) mechanism} is a function $\phi :\mathcal{P}%
\mathbf{\rightarrow }A(\mathbf{X})$ that returns an allocation in $A(\mathbf{%
X})$ for each profile of preferences $P\in \mathcal{P}.$\medskip

Given $d\in D,$ we denote by $\phi _{d}(P)$ the (unique) contract in $%
\phi (P)$ involving $d$, if such a contract exists. Otherwise,  $\phi _{d}(P)=\emptyset$.
\end{definition}

As previously mentioned, we assume that each hospital $h\in H$ has a
(fixed) antisymmetric, transitive and complete preference relation over $%
\mathbf{X}_{h} $, which is denoted by $P _{h}.$ The preference profile for hospitals will be denoted by $P_H :=(P _{h})_{h\in
H} $.

Let $Y\in A(\mathbf{X})$ be an allocation and $P\in \mathcal{P}$.  $Y$ is \textbf{individually rational} if  $Y_i R_i\emptyset$ for all $i \in D\cup H$, meaning that no agent gets an unacceptable contract. The allocation $Y$ is blocked by a contract $x\in \mathbf{X}\setminus Y$ if $x P_{x_{D}} Y_{x_D}$ and  $x P_{x_{H}} Y_{x_H}$ , i.e., if both involved agents prefer the contract $x$ to their assignment in $Y$.  The allocation $Y$ is \textbf{stable} if it is individually rational and is not blocked by any contract. 

\cite{hatfield2005matching} showed that the set of  stable allocations is always non-empty and forms a lattice, with respective unanimously preferred stable allocations for each side of the market at the extremes. 
Among the stable outcomes, we focus on the quantile stable allocations defined in \cite{chen2016median} and studied in \cite{chen2021quantile}, \cite{fernandez2020deferred}, and others.  The extremal quantile stable allocations are the doctor-optimal and the hospital-optimal ones. 

Take a $P\in \mathcal{P}$ and assume that $k$ is the number of stable allocations under $P$ (and $P_H$). Suppose that $\{X  ^1 , . . . , X  ^k\}$ is the set of all stable allocations in a given market. For each doctor $d$, consider the set of contracts that it signs  in such stable allocations: $\{X_d^1 , . . . , X_d  ^k\}$. Reorder these sets in a decreasing way according to doctor d's preferences,  ${X^{(1)}} _d , . . . ,{X^{(k)}}_d$, so that  ${X^{(1)}} _d$ coincides with the optimal stable assignment for $d$  and ${X ^{(k)}} _d$ coincides with the worst stable assignment for $d$.  Now, for $j=1,...,k$ we define $X^{(j)}:=\bigcup_{d\in D} {X ^{(j)}} _d$ , i.e., the set of each doctor's j-th best stable outcome.  
 \cite{chen2016median} showed that $X^{(j)}$ is a well-defined stable allocation for $j=1,...,k$ .  It is called the\textit{ j-th quantile stable allocation} for doctors. \footnote{Quantile stable allocations are also called \emph{generalized median-stable} allocations in \cite{chen2016median}.} 
 Considering quantile stable allocations,  
\cite{chen2014ranking,chen2016manipulability, chen2021quantile} introduce the next class mechanisms which select a quantile stable allocation respecting a "level" ($q$) of compromising between both side of the market.  

\begin{definition}
For each $q\in [0,1]$, the \textbf{$q$-quantile stable mechanism} $\phi ^{q}:\mathcal{P}%
\rightarrow A(\mathbf{X})$ selects the $\lceil kq\rceil $-th quantile allocation at each $P\in \mathcal{P} $, where $k$ is the number of stable allocations under $P$.\footnote{%
Here $\lceil x\rceil $ denotes the lowest positive integer equal to or
larger than $x$ if $x>0$, and we take $\lceil 0\rceil=1$.}
\end{definition}

The extreme quantile stable mechanisms $\phi ^{0}$ and $\phi ^{1}$ are called \emph{doctor-proposal DA} and \emph{hospital-proposal DA} mechanisms, respectively. 


The concept of non-manipulability, also known as strategy-proofness, has been central to studying agents' strategic behavior. A doctor is said to manipulate a matching mechanism if a situation exists where it achieves a better outcome by reporting a preference different from its true one. Formally, 
given a mechanism $\phi :$ $\mathcal{P}\mathbf{\rightarrow }A(\mathbf{X})$ and $d\in D$ with true preference $P_{d}\in \mathcal{P}_{d}$, the preference $P_{d}^{\prime }\in \mathcal{P}_{d}$ is a \textbf{manipulation} of $\phi $ at $P_{d}$ if there is a (sub)profile $P_{-d}\in \mathcal{P}%
_{-d}$ such that 
\begin{equation}
\phi _{d}(P_{d}^{\prime },P_{-d})\text{ }P_{d}\text{ }\phi
_{d}(P_{d},P_{-d}).
\end{equation}%
A mechanism is \textbf{non-manipulable} if no manipulation is possible.

It is known that the doctor-proposal DA ($\phi ^{0}$) is the only stable (quantile stable) mechanism that is not manipulable. Then, any $q$-quantile stable mechanism with $q>0$ is susceptible to manipulation. In this paper, we analyze if such manipulations are obvious in the sense of \cite{troyan2020obvious}. Intuitively, a manipulation is considered ``obvious'' if it makes the agent strictly better off either in the worst-case or best-case scenario.

Given a mechanism $\phi :\mathcal{P}\mathbf{\rightarrow }A(\mathbf{X}),$ a doctor $%
d\in D$ and a preference $P_{d}\in \mathcal{P}_{d}$, we define the \textbf{%
option set} left open by $P_{d}$ at $\phi $ as 
\begin{equation*}
O^{\phi }(P_{d})=\{\phi _{d}(P_{d},P_{-d}):P_{-d}\in \mathcal{P}_{-d}\}.
\end{equation*}

Given an allocation $Y \subseteq \mathbf{X}$, let $W_{d}(P_{d},Y)$ 
and $C_{d}(P_{d},Y)$ 
represent the worst and the best contract in $Y_d$ according to preference $P_{d}$, respectively. 


\begin{definition}\label{defnom}
Let $\phi :\mathcal{P}\mathbf{\rightarrow }A(\mathbf{X})$ be a mechanism,  $%
d\in D$ with true preference  $P_{d}\in \mathcal{P}_{d}$, and let $P_{d}^{\prime }\in \mathcal{%
P}_{d}$ be a manipulation of $\phi $ at $P_{d}$. Then, $P_{d}^{\prime
}$ is an \textbf{obvious manipulation} if 

\begin{equation} 
W_{d}(P_{d},O^{\phi }(P_{d}^{\prime }))\ P_{d}\ W_{d}(P_{d},O^{\phi
}(P_{d})).  \label{1}
\end{equation}%
or 
\begin{equation}
C_{d} ( P_{d},O^{\phi }(P_{d}^{\prime })) \ P_{d}\ C_{d} (
P_{d},O^{\phi }(P_{d})) .  \label{11}
\end{equation}%
Mechanism $\phi$ is \textbf{not obviously manipulable (NOM)} if it does not admit obvious manipulations. 
\end{definition}

 \section{Result}\label{section main}

 \cite{troyan2020obvious} proved that any stable (quantile stable) mechanism is not obviously manipulable in a model without contracts. \cite{arribillaga2025obvious}, demonstrated that, in a context with contracts, the doctor-proposing DA Mechanism remains not obviously manipulable while hospital-proposing DA becomes obviously manipulable. 
 Completing previous results, next theorem states that the doctor-proposal DA is the only quantile stable mechanism that remains not obviously manipulable in a context with contracts. As doctor-proposal DA ($\phi ^{0}$) is not manipulable, it is NOM.

In the case of $q$-quantile stable mechanisms with $q>0$, a doctor may falsely declare a particular “acceptable” contract as “unacceptable”  leading it to be rejected when proposed by the corresponding hospital. As a result, the hospital may offer more favorable terms to the same doctor in a subsequent round.  
Such manipulations could include only contracts between the manipulating doctor and the involved hospital, making these manipulations obvious.

\begin{theorem}
\label{quantile}A $q$-quantile stable rule $\phi ^{q}:\mathcal{P}\rightarrow
A(\mathbf{X})$ is NOM if and only if $q=0$, \textit{i.e.}, $\phi ^{q}$ is the doctor-proposal DA.
\end{theorem}

\noindent \begin{proof}
  As $\phi ^{0}$ is not manipulable, it is NOM. Now assume that $q\in (0,1],$ we will show that a particular market
exists where a doctor $d\in D$ has an obvious manipulation at $\phi ^{q}.$
Let $D=\{d_{1},d_{2}\}$, $H=\{h_{1},h_{2}\}$ be the sets of doctors and
hospitals, respectively. Let $k$ be a positive integer such that $\lceil
kq\rceil =2$ (observe that $k\geq2$)  and let $\mathbf{X}=\{x^{1},x^{2},\ldots,x^{k},w\}$ be a set of
contracts such that $x_{D}^{t}=d_{1},$ $x_{H}^{t}=h_{1}$ for each $t=1,\ldots,k$
and $w_{D}=d_{2},$ $w_{H}=h_{2}.$ Assume that $\succ
_{h_{1}}=x^{k},x^{k-1},\ldots, x^{1}$ and $\succ _{h_{2}}=w.$ Let $P_{2}$ be an
arbitrary preference in $\mathcal{P}_{d_{2}}.$ Let $P_{1}\in \mathcal{P}%
_{d_{1}}$ such that $P_{1}=x^{1},x^{2},\ldots, x^{k}.$ Observe that the set of
all stable allocations under $P$ is $\cup_{t=1\cdots k}\{\{x^{t},w\}\}$
if $wP_{2}\varnothing $ and $\cup_{t=1\cdots k}\{\{x^{j}\}\}$
if $\varnothing P_{2}w.$ As $\lceil kq\rceil =2,$ $\phi
_{d_{1}}^{q}(P)=x^{2}.$ Now, let $P_{1}^{\prime }\in \mathcal{P}_{d_{1}}$
such that $P_{1}^{\prime }=x^{1}.$ Then, the set of all stable allocations
under $(P_{1}^{\prime },P_{2})$ is $\{\{x^{1},w\}\}$ if $%
wP_{2}\varnothing $, and $\{\{x^{1}\}\}$ if $\varnothing P_{2}w.$ Hence $\phi
_{d_{1}}^{q}(P_{1}^{\prime },P_{2})=x^{1}.$ Therefore, $\phi
_{d_{1}}^{q}(P_{1}^{\prime },P_{2})=x^{1}P_{1}x^{2}=\phi
_{d_{1}}^{q}(P_{1},P_{2})$ for any $P_{2}\in \mathcal{P}_{d_{2}}.$ So, $%
P_{1}^{\prime }$ is an obvious manipulation of $\phi ^{q}$ at $P_{1}.$
\end{proof}

\begin{remark}
    Other stable mechanisms that also seek a compromise between both sides of the market are known as \textbf{interior-stable} mechanisms \citep{fernandez2020deferred}. These mechanisms aim to select a stable allocation that is neither doctor-optimal nor hospital-optimal, when such an option exists. It is straightforward to adapt our proof of Theorem \ref{quantile} to demonstrate that no interior-stable mechanism is NOM.     
\end{remark}

In some respects, our theorem aligns with the findings of \cite{chen2014ranking, chen2016manipulability}). These papers demonstrated that a quantile stable (stable) mechanism becomes "more manipulable" as it diverges from the doctor-proposing DA.  Another relaxation of strategy-proofness is the notion of regret-free truth-telling introduced by \cite{fernandez2020deferred}. This notion considers \emph{both} sides of the market and assumes that agents gain information about others' preferences by observing the rule's outcomes. It is proven that the doctor-proposing DA and the hospital-proposing DA are the only regret-free truth-telling mechanisms within the class of quantile stable mechanisms. Compared to the results in these papers, our findings further reinforce the idea that the doctor-proposing DA remains the strongest candidate among quantile stable mechanisms, even when considering weakened forms of strategy-proofness. They also highlight that the hospital-proposing DA would be susceptible to obvious manipulations. The comparative analyses with these works are limited. Our focus is on a specific weakness of strategy-proofness for one side of the market, whereas \cite{chen2014ranking} establishes a criterion to evaluate when a mechanism becomes "more manipulable for doctors" and \cite{fernandez2020deferred} examines a relaxation of strategy-proofness for agents on both sides of the market.

\bibliographystyle{ecta}
\bibliography{biblio}

\begin{thebibliography}{22}
\newcommand{\enquote}[1]{``#1''}
\expandafter\ifx\csname natexlab\endcsname\relax\def\natexlab#1{#1}\fi

\bibitem[\protect\citeauthoryear{Arribillaga and Bonifacio}{Arribillaga and Bonifacio}{2024}]{arribillaga2024obvious}
\textsc{Arribillaga, R.~P. and A.~G. Bonifacio} (2024): \enquote{Obvious manipulations of tops-only voting rules,} \emph{Games and Economic Behavior}, 143, 12--24.

\bibitem[\protect\citeauthoryear{Arribillaga and Bonifacio}{Arribillaga and Bonifacio}{2025}]{arribillaga2025not}
---\hspace{-.1pt}---\hspace{-.1pt}--- (2025): \enquote{Not obviously manipulable allotment rules,} \emph{Economic Theory}, 1--26.

\bibitem[\protect\citeauthoryear{Arribillaga and Pepa~Risma}{Arribillaga and Pepa~Risma}{2025}]{arribillaga2025obvious}
\textsc{Arribillaga, R.~P. and E.~Pepa~Risma} (2025): \enquote{Obvious Manipulations in Matching with and without Contracts,} \emph{Forthcoming in Games and Economic Behavior}.

\bibitem[\protect\citeauthoryear{Aziz and Lam}{Aziz and Lam}{2021}]{aziz2021obvious}
\textsc{Aziz, H. and A.~Lam} (2021): \enquote{Obvious manipulability of voting rules,} in \emph{International Conference on Algorithmic Decision Theory}, Springer, 179--193.

\bibitem[\protect\citeauthoryear{Chen, Egesdal, Pycia, and Yenmez}{Chen et~al.}{2014}]{chen2014ranking}
\textsc{Chen, P., M.~Egesdal, M.~Pycia, and M.~B. Yenmez} (2014): \enquote{Ranking by manipulability and quantile stable mechanisms,} \emph{Available at SSRN 2194458}.

\bibitem[\protect\citeauthoryear{Chen, Egesdal, Pycia, and Yenmez}{Chen et~al.}{2016{\natexlab{a}}}]{chen2016manipulability}
---\hspace{-.1pt}---\hspace{-.1pt}--- (2016{\natexlab{a}}): \enquote{Manipulability of stable mechanisms,} \emph{American Economic Journal: Microeconomics}, 8, 202--214.

\bibitem[\protect\citeauthoryear{Chen, Egesdal, Pycia, and Yenmez}{Chen et~al.}{2016{\natexlab{b}}}]{chen2016median}
---\hspace{-.1pt}---\hspace{-.1pt}--- (2016{\natexlab{b}}): \enquote{Median stable matchings in two-sided markets,} \emph{Games and Economic Behavior}, 97, 64--69.

\bibitem[\protect\citeauthoryear{Chen, Egesdal, Pycia, and Yenmez}{Chen et~al.}{2021}]{chen2021quantile}
---\hspace{-.1pt}---\hspace{-.1pt}--- (2021): \enquote{Quantile stable mechanisms,} \emph{Games}, 12, 43.

\bibitem[\protect\citeauthoryear{Dubins and Freedman}{Dubins and Freedman}{1981}]{dubins1981machiavelli}
\textsc{Dubins, L.~E. and D.~A. Freedman} (1981): \enquote{Machiavelli and the Gale-Shapley algorithm,} \emph{The American Mathematical Monthly}, 88, 485--494.

\bibitem[\protect\citeauthoryear{Echenique, Robinson-Cort{\'e}s, and Yariv}{Echenique et~al.}{2024}]{echenique2024experimental}
\textsc{Echenique, F., A.~Robinson-Cort{\'e}s, and L.~Yariv} (2024): \enquote{An experimental study of decentralized matching,} \emph{arXiv preprint arXiv:2401.10872}.

\bibitem[\protect\citeauthoryear{Fernandez}{Fernandez}{2020}]{fernandez2020deferred}
\textsc{Fernandez, M.~A.} (2020): \enquote{Deferred acceptance and regret-free truth-telling,} Working Paper.

\bibitem[\protect\citeauthoryear{Gale and Shapley}{Gale and Shapley}{1962}]{gale1962college}
\textsc{Gale, D. and L.~Shapley} (1962): \enquote{College admissions and the stability of marriage,} \emph{The American Mathematical Monthly}, 69, 9--15.

\bibitem[\protect\citeauthoryear{Hatfield and Milgrom}{Hatfield and Milgrom}{2005}]{hatfield2005matching}
\textsc{Hatfield, J. and P.~Milgrom} (2005): \enquote{Matching with contracts,} \emph{American Economic Review}, 95, 913--935.

\bibitem[\protect\citeauthoryear{Hirata and Kasuya}{Hirata and Kasuya}{2017}]{hirata2017stable}
\textsc{Hirata, D. and Y.~Kasuya} (2017): \enquote{On stable and strategy-proof rules in matching markets with contracts,} \emph{Journal of Economic Theory}, 168, 27--43.

\bibitem[\protect\citeauthoryear{Iwase}{Iwase}{2022}]{iwase2022equivalence}
\textsc{Iwase, Y.} (2022): \enquote{Equivalence theorem in matching with contracts,} \emph{Review of Economic Design}, 1--9.

\bibitem[\protect\citeauthoryear{Ortega and Segal-Halevi}{Ortega and Segal-Halevi}{2022}]{ortega2022obvious}
\textsc{Ortega, J. and E.~Segal-Halevi} (2022): \enquote{Obvious manipulations in cake-cutting,} \emph{Social Choice and Welfare}, 1--20.

\bibitem[\protect\citeauthoryear{Psomas and Verma}{Psomas and Verma}{2022}]{psomas2022fair}
\textsc{Psomas, A. and P.~Verma} (2022): \enquote{Fair and efficient allocations without obvious manipulations,} \emph{arXiv preprint arXiv:2206.11143}.

\bibitem[\protect\citeauthoryear{Roth}{Roth}{1982}]{roth1982economics}
\textsc{Roth, A.~E.} (1982): \enquote{The economics of matching: Stability and incentives,} \emph{Mathematics of operations research}, 7, 617--628.

\bibitem[\protect\citeauthoryear{Schwarz and Yenmez}{Schwarz and Yenmez}{2009}]{schwarz2009median}
\textsc{Schwarz, M. and M.~B. Yenmez} (2009): \enquote{Median stable matching for markets with wages,} \emph{Available at SSRN 1031277}.

\bibitem[\protect\citeauthoryear{Teo and Sethuraman}{Teo and Sethuraman}{1998}]{teo1998geometry}
\textsc{Teo, C.-P. and J.~Sethuraman} (1998): \enquote{The geometry of fractional stable matchings and its applications,} \emph{Mathematics of Operations Research}, 23, 874--891.

\bibitem[\protect\citeauthoryear{Troyan, Delacr{\'e}taz, and Kloosterman}{Troyan et~al.}{2020}]{troyan2020essentially}
\textsc{Troyan, P., D.~Delacr{\'e}taz, and A.~Kloosterman} (2020): \enquote{Essentially stable matchings,} \emph{Games and Economic Behavior}, 120, 370--390.

\bibitem[\protect\citeauthoryear{Troyan and Morrill}{Troyan and Morrill}{2020}]{troyan2020obvious}
\textsc{Troyan, P. and T.~Morrill} (2020): \enquote{Obvious manipulations,} \emph{Journal of Economic Theory}, 185, 104970.

\end{thebibliography}

\end{document}